\documentclass[final,5p,times,twocolumn]{elsarticle}
\usepackage{amsfonts}  
\usepackage{amsmath}
\usepackage[T1]{fontenc}
\usepackage[utf8]{inputenc}

\usepackage{lineno}

\usepackage{graphicx}
\usepackage{dcolumn}
\usepackage{booktabs}

\usepackage[colorlinks=true,linktocpage=true]{hyperref}
\hypersetup{
    colorlinks = true,
    linkcolor = red,
    anchorcolor = red,
    citecolor = red,
    filecolor = red,
    urlcolor = red
    }
\usepackage{setspace}
\usepackage{comment}
\usepackage{algorithm}
\usepackage{algpseudocode}
\usepackage{mattens}
\usepackage{nicematrix}

\usepackage[utf8]{inputenc}
\usepackage[T1]{fontenc}
\usepackage{mathptmx}
\usepackage{etoolbox}
\usepackage{hyperref}
\usepackage{algpseudocode}
\usepackage{algorithm}
\usepackage{comment}
\usepackage{lipsum}
\usepackage{siunitx}
\usepackage{tabularx}
\newcolumntype{Y}{>{\centering\arraybackslash}X}

\usepackage[flushleft]{threeparttable}

\usepackage{avant}

\newcommand {\T}[1]{{\mathbf{#1}}}

\begin{document}

\begin{frontmatter}


\title{Extracting flow stress surfaces of pristine materials using deformation paths in MD simulations}

\author[1,3]{E. T. Dubois}

\author[1,2]{Paul Lafourcade\corref{cor1}}
\cortext[cor1]{Corresponding author}
\ead{paul.lafourcade@cea.fr}

\author[1,2]{J.-B. Maillet}

\address[1]{CEA, DAM, DIF, F-91297 Arpajon, France}
\address[2]{Université Paris-Saclay, LMCE, F-91680 Bruyères-le-Châtel, France}
\address[3]{CEA, DES, IRESNE, DEC, SESC, LM2C, Cadarache F-13108 Saint-Paul-Lez-Durance, France}

\date{\today}
\begin{abstract}
  Accurate simulation of deformation processes at the atomic scale is critical for predicting the mechanical response of materials and particularly the calculation of directional flow stresses. This work presents a method for applying arbitrary deformation paths in \texttt{LAMMPS} while adhering to its convention that supercell periodic vectors \( \mathbf{a} \), \( \mathbf{b} \) are aligned such that \( \mathbf{a} \) coincides with the x-axis and \( \mathbf{b} \) lies in the (x,y) plane. This method is particularly relevant for materials with low crystal symmetry and also for exploring non uniaxial deformations. The first step of the method consists in generating the simulation frame tensor’s time evolution upon any deformation, which may initially violate \texttt{LAMMPS} alignment constraints. This constraint is then overcome by the application of a rigid body rotation to realign the tensor with \texttt{LAMMPS}'s convention, ensuring valid periodic boundary conditions. The resulting lengths and tilt factors from the rotated tensor are expressed analytically using third-order polynomials and applied to the simulation cell using the \texttt{fix deform} command. The present approach versatility is validated with the calculation of directional flow stresses for various materials upon constant volume shear, tension and compression, demonstrating its effectiveness in simulations involving complex deformation scenarios and diverse crystal structures. The flow stress surface extracted from these simulations are finally analyzed as the fingerprint of all deformation mechanisms occurring in the material.
\end{abstract}

\end{frontmatter}

\section{\label{sec:intro}Introduction}
Molecular dynamics (MD) simulation methods have significantly improved since their inception in the 1950s, becoming a crucial tool for understanding the behavior of materials at the atomic scale. Initially employed for simple systems, MD has evolved through improvements in computational power and algorithms, enabling the exploration of complex molecular interactions as well as microstructural features.
A. Rahman was instrumental in this evolution, as he pioneered the use of MD simulations in the 1960s to study the properties of liquids and solids \cite{rahman1964corr}. His groundbreaking work laid the foundation for applying MD to investigate material behavior under various conditions, including stress-induced deformation through the so-called Parrinello-Rahman barostat for example \cite{parrinello1981poly, rahman1984molecular}. Rahman's contributions helped establish the relevance of atomic-level simulations in understanding phenomena at the meso- and macroscopic scales.
As the development of multiscale approaches in computational materials science often incorporates a molecular dynamics (MD) component \cite{van2020roadmap}, one of its primary challenges lies in determining which elements should be coarse-grained, as this choice directly impacts the accuracy and reliability of the models at upper scales. Additionally, establishing a common language between different scales is crucial for effective communication and integration of the results. For example, the interplay between density functional theory (DFT) and classical MD has led to the creation of machine learning interatomic potentials (MLIP). These potentials have been validated through comparisons on specific properties such as elastic properties or stacking fault energy surfaces. In this scenario, the common language consists of the quantities of interest —specific properties or metrics— that can be computed using both DFT and classical MD.
Emphasizing one-to-one comparisons between these scales is vital for validating upscale models, as it allows the scientific community to assess the consistency and accuracy of predictions across different resolutions. By refining the connection between atomistic simulations and larger-scale models, multiscale approaches can enhance our understanding of material behavior under various conditions, ultimately leading to more robust and predictive modeling techniques.

However, transitioning from MD to continuum mechanics presents its own set of challenges, primarily due to the complexities involved in correlating the behavior of individual atoms with continuum mesh elements usually orders of magnitude larger in size. This correspondence is not straightforward and requires careful consideration.
In the context of developing mechanical constitutive laws, a crucial question arises: what are the minimum requirements needed to effectively link these laws in the continuum with the fluctuations and irreversible displacements of atoms at the microscopic scale? Several studies~\cite{Cai_mse_2004, Gilbert_prb_2011, Queyreau_prb_2011, Lafourcade_prm_2019, Denoual_jap_2024, Bertin_mt_2022, Lafourcade_mecmat_2024} have explored the construction of constitutive laws that incorporate microscopic information, often focusing on specific loading directions and simple crystallographic structures, where slip planes are well-defined and above all, specifically oriented with respect to the user-defined simulation frame.
Discrete dislocation dynamics (DDD) simulations represent another widely used method for coarse-graining the mechanical response of materials and consist in modeling individual or collective dislocations behavior at the mesoscale~\cite{Zhou_am_2010, Bulatov_nat_1998, Devincre_mse_1997, Madec_prl_2002, Devincre_science_2008}. In addition, recent works have demonstrated that capturing the dynamics of dislocations through large-scale MD simulations was indeed possible~\cite{Zepeda2017,Zepeda2020,Bertin2020}. However, the time and length scales reached by MD simulations imply high strain-rate conditions, only relevant for the materials behavior under shock loading for example.
Additionally, machine learning methods have recently been employed to directly identify mobility laws of such dislocation networks based on the data obtained from MD simulations in conjunction with graph neural networks~\cite{bertin2024learning}. While this approach is valuable, it is primarily limited to understanding plastic deformation mechanisms and does not encompass other deformation processes, such as twinning, fracture or phase transformation. Therefore, a more comprehensive framework that includes various deformation mechanisms is essential for developing robust constitutive models that can accurately predict material behavior across different scales.

The initial step in coarse-graining the mechanical response of materials to deformation involves identifying the various deformation mechanisms that may occur for every possible direction of applied deformation or stress state, effectively mapping this out on the unit sphere. This analysis can be encapsulated in a single representation known as the critical flow stress surface, previously introduced to the community taking as an example TATB (1,3,5-triamino-2,4,6-trinitrobenzene), a highly anisotropic energetic molecular crystal \cite{Lafourcade_jpcc_2018}. This surface serves as a comprehensive overview of all the deformation mechanisms that can be activated under different loading conditions, functioning as a unique fingerprint for each material's response to deformation. By capturing the relationship between the applied strain and the corresponding deformation mechanisms, the critical flow stress surface fully characterizes the material's behavior for a given loading. We argue that this surface is a crucial element in understanding crystal deformation, as it provides valuable insights into materials mechanical behavior. By integrating this critical flow stress surface into multiscale modeling efforts, one can enhance the understanding of deformation processes and incorporate lower scale physics into continuum models~\cite{Lafourcade_jap_2024}.

For a given type of loading —whether a traction, compression, shear, or a combination of these— the critical flow stress surface is constructed by applying a specific deformation path to the reference strain-free MD simulation cell. This process can be systematically repeated for any loading direction, although crystal symmetries may be used to significantly reduce the total number of required simulations. 
One challenge that arises in this context is the stringent constraints imposed by MD engines such as \texttt{LAMMPS} \cite{Thompson2022}, which limit the types of deformations that can be applied to a supercell due to additional constraints on the simulation frame in the Cartesian space. To address the latter, we propose a method that involves prescribing deformation paths that continuously adapt the shape of the supercell to comply with \texttt{LAMMPS} requirements. This is made possible by pre-computing the simulation frame time evolution and deriving the appropriate rigid-body rotation that allows the simulation frame to be in line with the \texttt{LAMMPS} definition. As \texttt{LAMMPS} is a widely used MD code in the scientific community, the aim of the present work is to propose a generic deformation-controlled MD framework available to all researchers interested in the mechanical response of materials and their multiscale modeling. Indeed, allowing deformation-controlled atomistic simulations can provide useful insights into the accuracy of continuum models by allowing one-to-one comparison across scales. 
This innovative approach enables us to effectively simulate constant strain rate deformations of materials in any arbitrary direction for various deformation types, thus providing a more flexible framework for exploring the mechanical responses of materials using large scale MD simulations. This method not only overcomes the limitations of the existing software but also opens new avenues for studying complex deformation mechanisms across different loading conditions.

In the present study, we focus on three distinct materials -namely graphite, silicon, tantalum- and extract their critical flow stress surfaces each corresponding to a different loading type at constant strain rate. By analyzing the shape of these surfaces, we demonstrate how to gain valuable insights into the underlying deformation mechanisms that govern each material's response to directional loading.
The resulting surfaces clearly illustrate the anisotropic nature of the mechanical behavior exhibited by each material, emphasizing how different loading directions can lead to varying responses. Finally we believe that these critical flow stress surfaces possess the potential to uncover the complexities and richness of materials mechanics. By providing a comprehensive view of how materials behave under various loading scenarios, they can inform more accurate predictive models and enhance our ability to tailor materials for specific applications. Finally, the developed software is entitled DEformation Paths for MOlecular Dynamics (\texttt{DEPMOD}) and allows for the generation of deformation paths module files suitable for both \texttt{LAMMPS} and \texttt{exaNBody}, the N-body HPC platform developed at CEA \cite{Carrard_2024}. The manuscript is organized as follows: we first describe the methods in Section~\ref{sec:theor} in which we clarify \texttt{LAMMPS} constraints and present \texttt{DEPMOD} in details. Section~\ref{sec:results} is then dedicated to the demonstration of \texttt{DEPMOD} usage on various materials under different loadings. We finally conclude and offer some potential evolution/perspectives in Section~\ref{sec:ccls}.

\section{\label{sec:theor}Methods}
In the present work, emphasis is placed on the application of macroscopic deformation paths to 3D-periodic molecular dynamics (MD) simulation cells. This framework was recently designed in the \texttt{exaNBody} code~\cite{Lafourcade_jpcc_2018, Carrard_2024, Lafourcade_mecmat_2024} to prescribe time-dependent deformations to MD and discrete element method (DEM) simulations. This time interpolation of deformation gradient tensors is made possible by built-in functions fully integrated in the \texttt{exaNBody} code. In addition, the core design of \texttt{exaNBody} does not impose any restrictions on the periodic vectors of the simulation cell, leaving any possibilities for applied deformations, in opposition to \texttt{LAMMPS} where the initial MD simulation frame tensor $\T{H_0}$ must be defined as follows:
\begin{equation}
    \T{H_0}=\begin{pmatrix} \T{a}_0 & \T{b}_0 & \T{c}_0 \end{pmatrix}=\begin{pmatrix} l_x^0 & xy^0 & xz^0 \\ 0 & l_y^0 & yz^0 \\ 0 & 0 & l_z^0\end{pmatrix}
\end{equation}
with $\T{a}$, $\T{b}$ and $\T{c}$ the MD simulation cell periodic vectors. In \texttt{LAMMPS}, $\T{a}$ and $\T{b}$ vectors must be parallel to the x-axis and contained in the (x,y) plane respectively, at any time during a MD trajectory. The current frame tensor is fully defined through $6$ variables separated in $3$ diagonal components ($l_x$, $l_y$, $l_z$) and $3$ tilt components ($yz$, $xz$, $xy$). In order to apply dynamic deformation to the MD simulation cell, \texttt{LAMMPS} offers the ability to evolve these $6$ variables through the \texttt{fix deform} command. The latter is however limited to deformations that satisfy the constraints on $\T{a}$ and $\T{b}$ vectors. 
A discussion on \texttt{LAMMPS} restriction can be found in its documentation in the \textbf{How to triclinic}\footnote{\url{https://docs.lammps.org/Howto_triclinic.html}} section. \texttt{LAMMPS} systematically uses the restricted triclinic framework internally in opposition to the general triclinic one allowed by \texttt{exaNBody}. The main idea of this work is therefore to lift the lock on mechanical deformation applicable to simulation boxes in \texttt{LAMMPS} by providing users with a feature similar to that of \texttt{exaNBody} and its \texttt{exaStamp} MD application.

Taking $\T{H_0}$ as the reference frame tensor, in the general case any current frame tensor $\T{H}(t)$ can be defined using the time-dependent deformation gradient tensor $\T{F}$, as usually done in continuum mechanics:
\begin{equation}
    \T{H}(t)=\begin{pmatrix} \T{a} & \T{b} & \T{c} \end{pmatrix}=\T{F}(t)\T{H_0}
    \label{eq:tdepstrain}
\end{equation}
As it is, the deformation can take any forms through $\T{F}(t)$, including non-proportional strain. Such deformation paths applied to large atomistic samples can help trigger material-specific deformation mechanisms and study their affinity to loading direction~\cite{Lafourcade_jpcc_2018, Lafourcade_prm_2019, Lafourcade_jap_2024}. However, applying such unconstrained deformation gradient tensor to the reference frame can, in some case, violate the \texttt{LAMMPS} convention explained before. The present work aims at providing a general method that uses built-in \texttt{LAMMPS} functions to apply any deformation gradient tensors to MD simulations cells and overcome these constraints. 

\subsection{\label{subsec:lammps} Overcoming \texttt{LAMMPS} constraints}
Given the reference frame tensor $\T{H_0}$, the application of a deformation gradient tensor $\T{F}(t)$ defines its time evolution $\T{H}(t)$ as explained in detail before. While arbitrary deformation can lead to a violation of \texttt{LAMMPS} convention, the axes of periodicity are preserved at any time. One way to overcome these constraints is to perform a pre-computation of the current frame tensor time evolution and calculate, for each time discretization, the corresponding rigid body rotation that allows to bring back $\T{a}$ and $\T{b}$ vectors into the ($x, y$) plane:
\begin{equation}
    \T{\tilde{H}}(t)=\begin{pmatrix} \T{\tilde{a}} & \T{\tilde{b}} & \T{\tilde{c}} \end{pmatrix}=\T{Q}(t)\T{H}(t)
    \label{eq:rotation}
\end{equation}
with $\T{Q(t)}$ the instantaneous rigid body rotation and $\T{\tilde{H}}(t)$ the new frame tensor time evolution that respects the constraints. It is to be noted that $\T{Q}(t)$ is built such that $\T{\tilde{a}}$ is aligned with the x-axis and $\T{\tilde{b}}$ lies in the (x,y) plane, meaning that it evolves with time, depending on the applied deformation $\T{F}(t)$. The condition $\T{\tilde{H}}(0)=\T{H_0}$ is considered as an additional constraint. It is important to note that the true time evolution of the frame tensor $\T{H}(t)$ is entirely generated once and for all before calculating the rigid body rotation through $\T{Q}(t)$. Then and only then, the time evolution of the \texttt{LAMMPS}-compatible frame tensor $\T{\tilde{H}}(t)$ is used to identify the time evolution of the $6$ independent parameters ($l_x$, $l_y$, $l_z$, $yz$, $xz$, $xy$). This means that depending on the user-defined time-dependent deformation gradient tensor, the time evolution of the frame tensor and its rigid body motion to transform it from general triclinic to restricted triclinic are known in advance and computed only once.

The MD simulation cell is finally dynamically strained through the \texttt{fix deform} \footnote{\url{https://docs.lammps.org/fix_deform.html}}, as implemented in \texttt{LAMMPS}, by providing analytical functions that describe the $6$ parameters variations time evolution $\Delta L$(t) as well as their respective variation rate $\dot{\Delta L}$(t). In the present work, the analytical functions are chosen to be $3^\mathrm{rd}$ order polynomials of the form:
\begin{equation}
    \Delta L(t) = At +Bt^2+Ct^3+D
    \label{eq:deltaL}
\end{equation}
with $L$ any of the ($l_x$, $l_y$, $l_z$, $yz$, $xz$, $xy$) parameters. Whether the deformation is applied under true or engineering constant strain-rate, the MD supercell lattice parameters can evolve linearly or exponentially. In order for the process to work any scenario, $3^\mathrm{rd}$ order polynomials seem a reasonable choice. These $6$ polynomials are independently fitted to the analytical evolution of $\T{\tilde{H}}(t)$, calculated through Equations~\ref{eq:tdepstrain} and~\ref{eq:rotation}. The rate of change is trivially derived from Equation~\ref{eq:deltaL}. Note that in Equation~\ref{eq:deltaL}, $D$ is not fitted nor fixed to $0$ but rather used as an offset during the restart of a simulation, allowing for a continuous deformation of the MD simulation cell across restarts. 

\subsection{\label{subsec:depmod} \texttt{DEPMOD}: DEformation Paths for MOlecular Dynamics}
The entire process described above is summarized in Algorithm~\ref{algo1} and analytical computations of $\T{F}(t)$, $\T{H}(t)$, $\T{Q}(t)$ and $\T{\tilde{H}}(t)$ are performed with the \texttt{DEPMOD} (\textbf{DE}formation \textbf{P}aths for \textbf{MO}lecular \textbf{D}ynamics) python package presented hereafter. \texttt{DEPMOD} was developed in the context of this work and is freely available at \href{https://github.com/lafourcadep/depmod}{https://github.com/lafourcadep/depmod}. This pre-process tool is dedicated to the generation of deformation paths for classical simulations performed with either \texttt{exaNBody} or \text{LAMMPS}. These deformation paths aim at defining the time evolution of the periodic boundary conditions vectors of the simulation. 
To generate a specific deformation path, \texttt{DEPMOD} takes the following general arguments as input:
\begin{itemize}
    \item Initial frame tensor $\T{H_0}$, either from 3 lattice vectors or by providing an initial simulation cell file from which the frame tensor is automatically extracted,
    \item A target strain-rate $\dot{\gamma}$, 
    \item A time interval $[t_\mathrm{min},t_\mathrm{max}]$ where $t_\mathrm{min}$ specifies the time at which the deformation starts and $t_\mathrm{max}$ the final time of the non-equilibrium trajectory,
    \item Two integers $N$ and $k$ where $N$ is the number of points used to fit analytical functions to the frame tensor components time evolution and $k$ is the number of sampling points for visualization and verification purposes,  
    \item A deformation mode, e.g. traction, compression or pure shear.
\end{itemize}

For the last item, different types of constant strain-rate deformations are pre-defined in \texttt{DEPMOD} and are explained in detail below. Starting with $\T{F}(t=0)=\T{I}$, i.e. the identity matrix, the increment of deformation corresponding to a time increment $\Delta t$ uses the usual relation between velocity gradient $\T{L}$ and strain gradient, namely $\dot{\T{F}}\cdot\T{F}^{-1}=\T{L}$. In its incremental formulation, the time evolved deformation gradient tensor reads:
\begin{equation}
    \T{F}(t+\Delta t)=\T{F}(t)+  \T{L} \cdot \T{F}(t) \Delta t
\end{equation}
where $\T{L}$ is chosen constant in the present case.

\subsubsection{\label{subsubsec:traction}Traction/compression}
For a pure traction or compression, the velocity gradient tensor is defined as follows:
\begin{equation}
    \T{L}=\dot{\gamma} \cdot \T{\tilde{m}} \otimes \T{\tilde{m}},
\label{eq:Ftrac}    
\end{equation}
with $\T{\tilde{m}}$ the normalized traction/compression vector and $\dot{\gamma}$ the imposed strain-rate. Eventually, the sign of $\dot{\gamma}$ will determine whether a traction or a compression is applied to the MD simulation cell. In \texttt{DEPMOD}, $\T{m}$ can be specified using either the $\T{[hkl]}$ notation or two angles $(\theta,\phi)$ from which $\T{m}$ is deduced as follows:
\begin{equation}
    \T{m} = \begin{bmatrix} \sin \theta \cos \phi  \\ \sin \theta \sin \phi \\ \cos \theta \end{bmatrix}
\end{equation}
with $\theta$ the polar angle, i.e. the angle with respect to the polar z-axis and $\phi$ the azimutal angle corresponding the the angle from the initial meridian plane defined as the (x,z) plane, in spherical coordinates.

\subsubsection{\label{subsubsec:isoVtraction}Isochoric traction/compression}
In some cases, maximizing the shear stress perceived by the sample can lower the stress at nucleation of specific deformation mechanisms. This can be interesting when one tries to identify new mechanisms or enforce their activation. To do so, one can choose to apply an isochoric deformation, meaning that a transverse and proportional traction/compression is applied to the sample if the latter is subject to a compression/traction, respectively. In that case, the velocity gradient tensor reads:
\begin{equation}
    \T{L}=\dot{\gamma} \cdot \T{\tilde{m}} \otimes \T{\tilde{m}} + \frac{1}{\Delta t} \left( \frac{1}{\sqrt{1 + \Delta t \dot{\gamma}}}-1\right) \left( \T{I} - \T{\tilde{m}} \otimes \T{\tilde{m}}\right),
\label{eq:Fisovtrac}       
\end{equation}
where the right hand side ensures that a transverse deformation at an appropriate strain-rate is applied to maintain a constant volume during the deformation trajectory. Here also, the sign of $\dot{\gamma}$ fully determines whether a traction or a compression is applied along the direction defined by $\T{m}$, as for the pure traction/compression case. In \texttt{DEPMOD}, the $\T{m}$ direction is specified in the same way as the pure traction/compression case. However, the additional parameter \texttt{isoV} has to be provided to ensure the isochoric character of the generated deformation path.

\subsubsection{\label{subsec:pureshear}Pure shear}
Finally, the last scenario discussed in this paper is a pure shear trajectory. The corresponding velocity gradient reads:
\begin{equation}
    \T{L}=\dot{\gamma} \cdot \T{\tilde{m}} \otimes \T{\tilde{n}},
\label{eq:Fshear}       
\end{equation}
with $\T{\tilde{m}}$ and $\T{\tilde{n}}$ the normalized vectors that fully characterize the applied shear deformation. In that case, $\T{\tilde{m}}$ and $\T{\tilde{n}}$ correspond to the shear direction and shear habitat plane respectively. In \texttt{DEPMOD}, multiple scenarios can be used to define the shear direction/plane; 1) $\T{m}$ and $\T{n}$ can be directly specified as vectors using the $\T{[hkl]}$ notation. 2) Two angles $(\theta,\phi)$ are provided, from which both $\T{m}$ and $\T{n}$ are deduced using the following:
\begin{equation}
    \T{m} = \begin{bmatrix} \sin \theta \cos \phi  \\ \sin \theta \sin \phi \\ \cos \theta \end{bmatrix}
    \T{n} = \begin{bmatrix} \cos \theta \cos \phi  \\ \cos \theta \sin \phi \\  -\sin \theta \end{bmatrix}
\end{equation}
where in that case, the plane defined by $(\T{m},\T{n})$ is chosen to systematically contain the z-axis, for simplicity. In the first case, an additional parameter, namely \texttt{strict}, can be provided in order to test whether the two vectors defining the required shear deformation are orthogonal to each other.

\begin{table}[!t]
\small
\caption{\label{tab:depmod}\texttt{DEPMOD} Input parameters for different deformation modes.}
\centering
\vspace{0.1cm}
\begin{tabularx}{\linewidth}{YYY}
\hline
\hline
Notation & Type & Units\\
\hline
\hline
\multicolumn{3}{c}{Parameters common to all modes}\\
\hline
$\T{H}_0$ & Mat3d or file & N/A \\
$\dot{\gamma}$ & double & s$^{-1}$ \\
t$_\mathrm{min}$ & double & s \\
t$_\mathrm{max}$ & double & s \\
N & int & N/A \\
k & int & N/A \\
\hline
\multicolumn{3}{c}{Parameters for traction/compression}\\
\hline
$\theta$ & double & deg \\
$\phi$ & double & deg \\
$\T{m}$ & Vec3d{} & N/A \\
\texttt{isoV} & bool & N/A \\
\hline
\multicolumn{3}{c}{Parameters for pure shear}\\
\hline
$\theta$ & double & deg \\
$\phi$ & double & deg \\
$\T{m}$ & Vec3d{} & N/A \\
$\T{n}$ & Vec3d{} & N/A \\
\texttt{strict} & bool & N/A \\
\hline
\hline
\end{tabularx}

\end{table}

\subsubsection{\label{subsec:exaStamp}\texttt{exaNBody} and \texttt{LAMMPS} framework}

To sum up, the entire set of parameters taken as input by \texttt{DEPMOD} are reported in Table~\ref{tab:depmod}. In addition, we provide the pseudocode defining the entire calculation chain in Algorithm~\ref{algo1}. In practice, the reference frame tensor is specified either by a 3x3 matrix or by providing an input file from which it is automatically extracted. Depending on the required deformation, associated directions $(\T{m}, \T{n})$ and strain-rate $\dot{\gamma}$, \texttt{DEPMOD} automatically generates the time evolution of the deformation gradient tensor $\T{F}(t)$, the true frame tensor $\T{H}(t)$, as well as the \texttt{LAMMPS}-compatible frame tensor $\T{\tilde{H}}(t)$ obtained through the rotation $\T{Q}(t)$.
\begin{algorithm}[H]
   \small
   \caption{\texttt{DEPMOD} workflow}
    \begin{algorithmic}
    \State{}
    \State{$\T{H}_0$ : Reference frame tensor}
    \State{$\dot{\gamma}$ : Strain-rate}
    \State{$t_\mathrm{min}$ : Deformation starting time}
    \State{$t_\mathrm{max}$ : Deformation final time}    
    \State{$N$ : Number of samples for polynomial fitting}
    \State{$k$ : Number of samples for output}
    \State{Set deformation type (traction/compression/shear)}
    \State{$(\T{m},\T{n})$ : Deformation direction(s)}
    \State{}
    \If{\texttt{exaNBody} output}
        \State{Evaluate $\T{F}(t)$ at $k$ times $\in$ $[t_\mathrm{min},t_\mathrm{max}]$}
        \State{Output \texttt{exaNBody} module file}
    \ElsIf{\texttt{LAMMPS} output}
        \State{Evaluate $\T{F}(t)$ and $\T{H}(t)$ at $N$ times $\in$ $[t_\mathrm{min},t_\mathrm{max}]$}
        \State{Calculate rotation matrix $\T{Q}(t)$ to satisfy constraints}
        \State{Calculate \texttt{LAMMPS}-compatible $\T{\tilde{H}}(t)$}
        \State{Fit $\T{\tilde{H}}(t)$ components to $3^\mathrm{rd}$-order polynomials}
        \State{Output \texttt{LAMMPS} module file}
    \EndIf
    \State{}
    \State{Evaluate $\T{F}(t)$, $\T{H}(t)$, $\T{Q}(t)$ and $\T{\tilde{H}}(t)$ at $k$ times $\in$ $[t_\mathrm{min},t_\mathrm{max}]$ and output .csv file for post-processing and graphical purposes}
    \end{algorithmic}
\label{algo1}
\end{algorithm}
Finally, \texttt{DEPMOD} outputs the file to be included in either the \texttt{exaNBody} or \texttt{LAMMPS} input deck. Since \texttt{exaNBody} already includes a formalism that allows to directly specify a tabular evolution of the deformation gradient tensor, $\T{H}(t)$, $\T{\tilde{H}}(t)$ and $\T{Q}(t)$ are not calculated but only the evaluation of $\T{F}(t)$ is performed. On the opposite, an extra step is performed to generate a \texttt{LAMMPS} compatible input deck by fitting the time-evolution of $\T{\tilde{H}}(t)$ components to $3^\mathrm{rd}$ order polynomials, allowing to pass them as variable to a \texttt{fix deform} command automatically generated by \texttt{DEPMOD}. What then differs between \texttt{exaNBody} and \texttt{LAMMPS} is only a time-dependent rigid body rotation.

To ensure that both frameworks lead to the same governing physics, we performed a toy simulation on a small BCC tantalum sample containing 2000 atoms. The sample was generated using Atomsk~\cite{atomsk} and atomic positions were disturbed by imposing a random displacement using a uniform distribution between -0.1 and +0.2 \AA. This sample, without any initial atomic velocities, was taken as an input of \texttt{DEPMOD} to generate appropriate deformation paths module files for both \texttt{LAMMPS} and \texttt{exaStamp} codes. The deformation path consisted in an isochoric traction along the $\T{[102]}$ crystallographic axis at a strain-rate $\dot{\gamma}=1\mathrm{e}^9$ s$^{-1}$ for a total of 100 ps. Starting with the exact same input sample, the latter is dynamically strained using both MD codes in the NVE ensemble with no additional thermostat or barostat, for the sake of comparison.

\begin{figure}[!h]
\centering
\includegraphics[width=\linewidth]{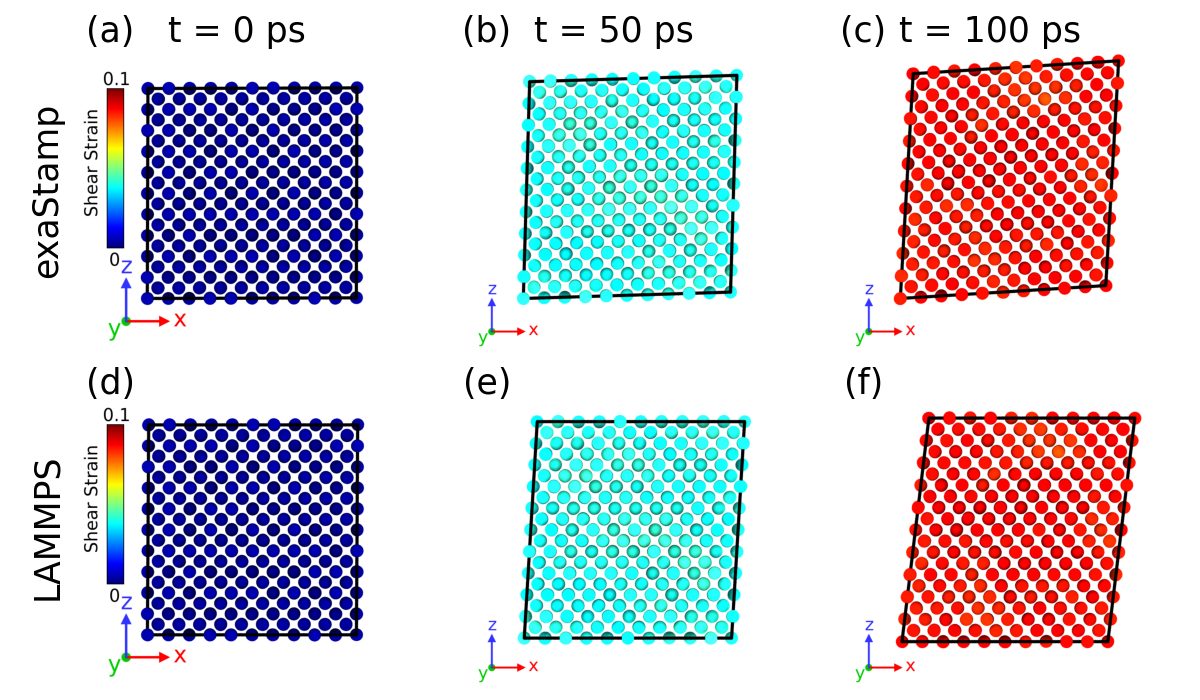}
\caption{\label{fig:fig1} Comparison between \texttt{exaStamp} (top) and \texttt{LAMMPS} (bottom) simulation cell evolution at 0, 50 and 100 ps corresponding to labels \textbf{(a),(d)}, \textbf{(b),(e)} and \textbf{(c),(f)} respectively.}
\end{figure}

To highlight the difference between \texttt{exaStamp} and \texttt{LAMMPS} formalisms w.r.t. deformation prescription to the simulation cell, snapshots of both samples at three different times are displayed in Figure~\ref{fig:fig1}, where the atoms are colored with respect to their local shear strain computed using \texttt{OVITO}~\cite{ovito_ref}. While the local value of the shear strain remains the same between both MD simulations, as well as the macroscopic deformation, it remains evident that both simulations evolve in a different laboratory frame, i.e. differing by a rigid-body rotation. One can also notice that each time corresponds to a different rotation matrix, as explained before. The pre-computation of such time-dependent rotation is the ultimate goal of the \texttt{DEPMOD} framework.

\begin{figure}[!h]
\centering
\includegraphics[width=0.8\linewidth]{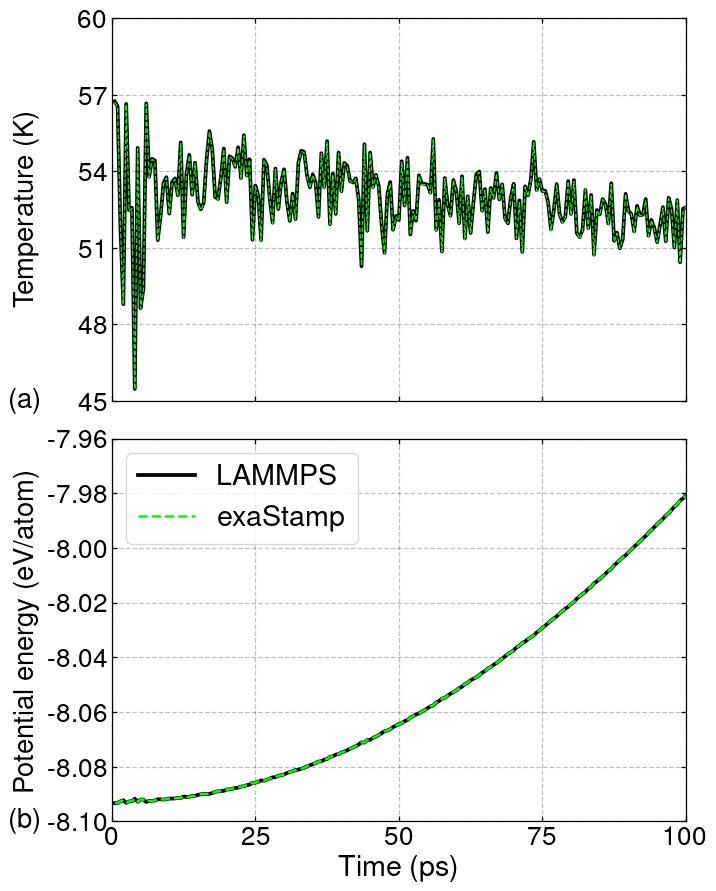}
\caption{\label{fig:toy_vars} Comparison of temperature (a) and potential energy (b) time evolution. \texttt{exaStamp} and \texttt{LAMMPS} results are displayed in continuous black and dashed green lines respectively.}
\end{figure}
In addition, both temperature and per-atom average potential energy evolution with time are displayed in Figure~\ref{fig:toy_vars}, showing that both sample follow the same non-equilibrium trajectory, but in a different frame as evidently represented with the samples snapshots in Figure~\ref{fig:fig1}. These demonstration case shows \texttt{DEPMOD}'s ability to pre-compute complex deformation paths for MD simulations using \texttt{LAMMPS}, while preserving the physical meaning of the applied strain. In the next section, we demonstrate the benefits of using \texttt{DEPMOD} to extract pristine materials flow stresses upon directional deformation, which can be useful to correlate the mechanical response of materials at the microscopic scale in the context of multiscale homogenization.

\section{\label{sec:results}Microscopic flow stress surfaces}
Three materials, i.e. hexagonal graphite (C), cubic diamond (c-dia) silicon (Si) and body-centered cubic (bcc) tantalum (Ta) were considered in the present work. All MD simulations were performed using \texttt{LAMMPS} to demonstrate the capabilities of the proposed framework. The three samples were first equilibrated in the NPT ensemble (using the aniso keyword) for 100 ps at 300 K, 0 GPa except for Si for which the temperature was set to 1000 K. The timestep for integrating the equations of motion was set to 1 fs for Si and Ta and 0.5 fs for C. Coupling constants for both thermostat and barostat were set to 100 fs and 1 ps, respectively. For computing interatomic interactions, the Airebo-morse~\cite{Oconnor_jcp_2015}, Embedded Atom Model (EAM)~\cite{Ravelo_aip_2012} and Stillinger-Weber~\cite{stillinger_prb_1985} interatomic potentials were respectively used for graphite, tantalum and silicon. After the NPT equilibration, the samples were thermalized for another 100 ps in the NVT ensemble at a fixed volume set as the average volume of the last 25 ps of the NPT trajectory. The final states of these NVT trajectories were taken as the starting configurations for the MD simulations involving deformation paths generated with \texttt{DEPMOD}. Properties of the different initial simulation cells are listed in Table~\ref{tab:mdsim}. It is to be noted that larger samples were considered for Ta because of the low computational cost of the potential used. 
\begin{table}[!h]
\caption{\label{tab:mdsim}Properties of the different MD initial simulation cells for the three different materials considered in this work.}
\vspace{0.1cm}
\resizebox{\linewidth}{!}{
  \begin{tabular}{ccrcccr}
    \hline
\hline    
Material & Potential & T (K) & L$_x$ (\AA) & L$_y$ (\AA) & L$_z$ (\AA) & N$_\mathrm{atoms}$\\
\hline
C & \cite{Oconnor_jcp_2015} & 300 & 154.88 & 150.84 & 160.10 & 442,368 \\
Si & \cite{stillinger_prb_1985} & 1000 & 261.54 & 261.54 & 261.54 & 884,736 \\
Ta & \cite{Ravelo_aip_2012} & 300 & 476.52 & 476.52 & 476.52 & 5,971,968 \\
\hline
\hline
\end{tabular}
}
\end{table}

Each material was subjected to a set of multiple deformation paths along different directions spanning the unit sphere, or a subset of it depending on the crystalline symmetry of each material. In each case, the strain-rate $\dot{\gamma}$ was set to $2\times10^8$ s$^{-1}$ and the trajectory ran until the appearance of a stress drop, i.e. corresponding to the nucleation of a given deformation mechanism. Spanning all orientations for a specific deformation allows to investigate the mechanical response of microscopic samples and its dependence on the loading direction. Specifically, and in particular for pristine samples where no initial defects are present, this methodology allows to compute the microscopic flow stress surface that in fact corresponds to the critical stress at nucleation of the first deformation mechanism. This was previously performed on TATB, an energetic molecular crystal, upon directional pure shear~\cite{Lafourcade_jpcc_2018}. It was found that the microscopic flow stress surface was essential to unveil the diversity of deformation mechanisms. In addition, the directional aspect of this surface usually correlates well with the crystalline symmetry of the studied sample. Finally, in the context of informing mesoscale models, such surface can help calibrate crystal plasticity models at the mesoscale and in particular the critical stresses for plastic activity on individual slip systems as was done for TATB single crystal~\cite{Lafourcade_jap_2024}. In the following, as different deformation types were considered for each material, the corresponding results are discussed in separate sections.

\subsection{\label{subsec:graphite}Graphite upon isochoric traction}
Isochoric traction was applied to graphite single crystal by defining the loading direction using spherical coordinates and the deformation gradient defined in Equation~\ref{eq:Fisovtrac}. The direction was defined using the polar angle $\theta$, i.e. the angle with respect to the polar z-axis as well as the azimutal angle $\phi$ corresponding the the angle from the initial meridian plane defined as the (x,z) plane. $(\theta,\phi)$ were both chosen between \ang{0} and \ang{90} with a \ang{15} step for $\phi$ and a \ang{30} for $\theta$, leading to a total of 28 independent deformation paths automatically generated using \texttt{DEPMOD}. This number was reduced to 22 since for $\theta=0$, any value for $\phi$ leads to a traction along the z-axis with a transverse compression independent to its value. Using the hexagonal crystallographic notation, directions $\T{[2\bar{1}\bar{1}0]}$, $\T{[\bar{1}2\bar{1}0]}$ and $\T{[0001]}$ were aligned to the x-, y- and z-axis respectively. This means that the graphitic layers were contained in the (x,y) plane with their normal parallel to the z-axis.
\begin{figure}[!h]
\centering
\includegraphics[width=0.9\linewidth]{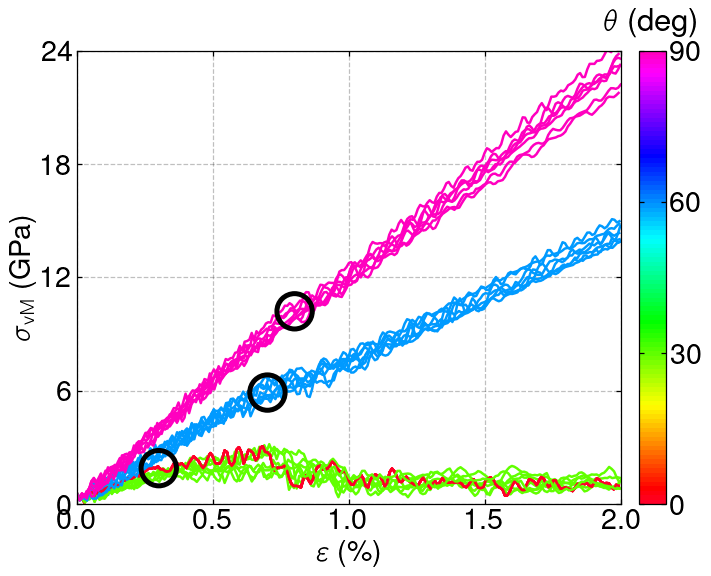}
\caption{\label{fig:curves_graphite} Stress-strain curves obtained for graphite single crystal under isochoric traction. Curves are colored according to the polar angle $\theta$ that partially defines the loading direction. Black circles indicate the area where the buckling instability nucleates.}
\end{figure}
\begin{figure*}[!t]
\centering
\includegraphics[width=\linewidth]{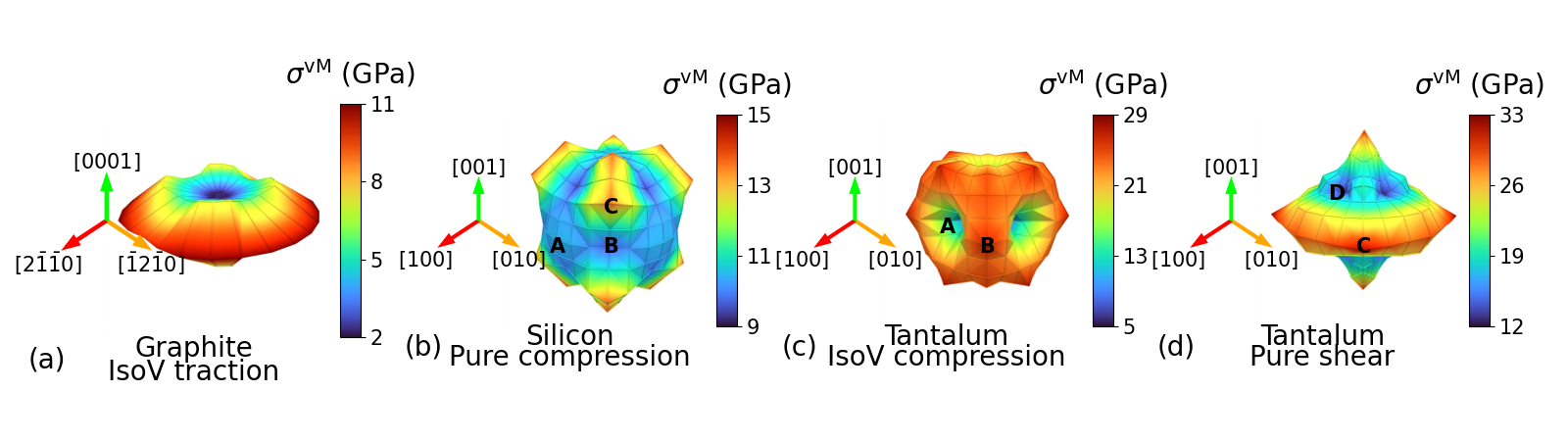}
\caption{\label{fig:flow_full} Critical shear stress surfaces for directional deformation of various pristine single crystals. \textbf{(a)} Graphite under isochoric traction. \textbf{(b)} Silicon under pure compression. \textbf{(c)} Tantalum under isochoric compression. \textbf{(d)} Tantalum under pure shear.}
\end{figure*}
Along the trajectories, the von Mises shear stress was calculated from the global stress tensor computed within the MD simulations while the shear strain was calculated in a Lagrangian formalism, i.e. with respect to the reference frame to which the deformation path was applied. Stress-strain curves are represented in Figure~\ref{fig:curves_graphite} and colored according to the value of $\theta$. A striking result is that curves are grouped with respect to their values of $\theta$ which can be attributed to the strong directional dependence of graphite mechanical response upon uniaxial traction. The greater the $\theta$ value, the stiffer the stress response, indicating a very large anisotropy in agreement with graphite single crystal elastic tensor. This is coherent with the longitudinal elastic constants of graphite with approximate values of 30 GPa and 1000 GPa along directions transverse to and within its basal plane~\cite{Polewczyk_cms_2023}. 
\begin{figure}[!b]
\centering
\includegraphics[width=0.75\linewidth]{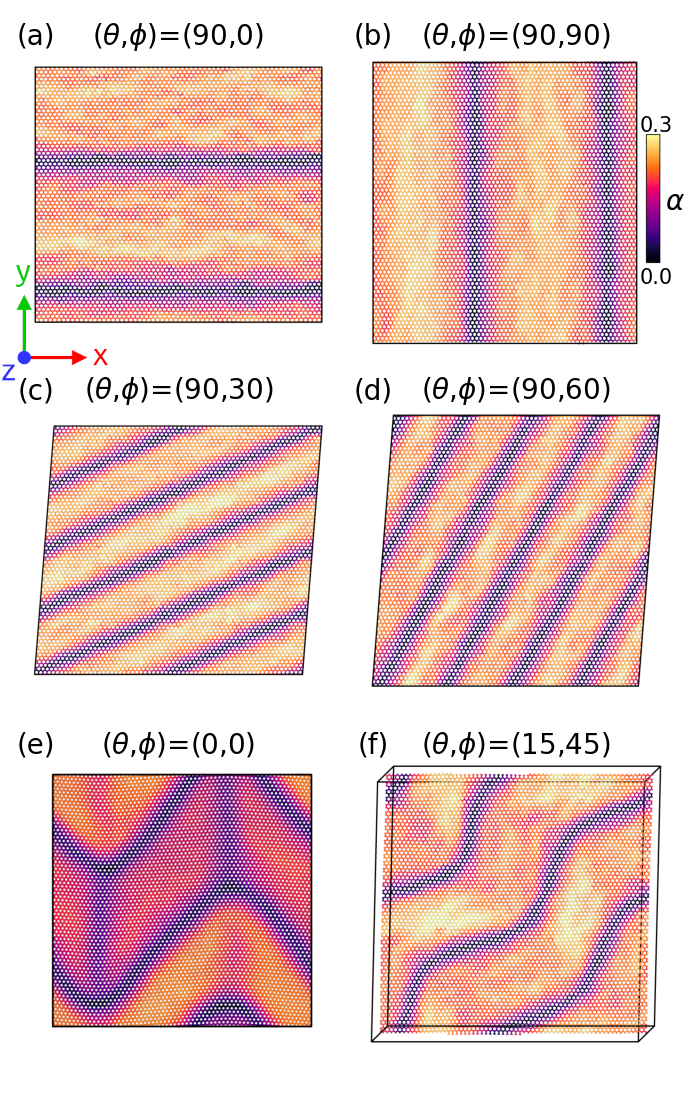}
\caption{\label{fig:ustructure_graphite} Buckling patterns observed in graphite single crystal under isochoric traction along various directions defined by $(\theta,\phi)$ in spherical coordinates.}
\end{figure}

Additionally, a decrease of the slope of the stress-strain curves as $\theta$ decreases is observed, i.e. as the traction direction shifts from the basal plane towards the z-axis. This change in stress-strain slope is attributed to a buckling instability triggered due to the isochoric character of the imposed deformation. Indeed, the volume-preserving transverse compression that compensates the uniaxial traction systematically induces a significant compression in a stiff direction, leading to the onset of elastic buckling, as previously observed in highly anisotropic and layered materials~\cite{Lafourcade_jpcc_2018, Lafourcade_prm_2019, Gruber_jmps_2024,barsoum4875627role}.
The von Mises shear stress at the onset of buckling is automatically extracted and used to build the critical shear stress surface of graphite, displayed in Figure~\ref{fig:flow_full}a. While the range in $(\theta,\phi)$ only allowed to actually compute 1/8 of the sphere, the full surface is reconstructed using reflections and symmetries of graphite. The resulting surface is in agreement with the transversely isotropic behavior of graphite and indicates a large difference in the critical stress value whether the traction is applied transversely or within the graphitic layers. On Figure~\ref{fig:curves_graphite}, black circles indicate the area where the buckling instability nucleates, which correlate to the value observed on the critical stress surface that do not depend on $\phi$ angle.
Finally, we display on Figure~\ref{fig:ustructure_graphite} the microstructural state after the critical stress has been reached, indicating the stabilization of a pattern induced by the elastic instability. 
In that figure, one single graphenic layer located in the middle of the sample in the reference frame was tracked along the trajectory for visualization with all other removed to avoid Moiré patterns. On that specific graphenic layer, bonds were created with a 1.9 {\AA} cutoff and atoms were removed. The colormap is based on the value of $\alpha=\sqrt{F^\mathrm{loc}_{xz} \cdot F^\mathrm{loc}_{xz} + F^\mathrm{loc}_{yz} \cdot F^\mathrm{loc}_{yz}}$ where $\T{F^{loc}}$ is the per-atom deformation gradient tensor calculated using \texttt{OVITO}~\cite{ovito_ref} on the all-atoms system. Per-particle data was subsequently transferred to the bonds for visualization. The darken area correspond to the interfaces between the buckled parts and correlates very well with the traction direction, i.e. they systematically align with the transverse direction associated with a compressive component of the deformation. For example, snapshots (a) and (b) correspond to traction along the x- and y-axis with a compression along the y- and x- axis respectively. In addition, snapshots (c) and (d) correspond to traction in the basal plane with $\phi$ equal to \ang{30} or \ang{60}, which lead to interfaces with the exact same angle. Snapshot (e) corresponds to a pure traction along the z-axis, implicating a traction along both x- and y-axis. In that case, the entire basal plane is subjected to a homogeneous compressive strain, leading to a 2D buckling pattern previously observed in graphite~\cite{Lafourcade_jap_2020} and TATB~\cite{Lafourcade_jpcc_2018} subjected to triaxial compression. Finally, snapshot (f) displays the resulting pattern for a traction along an axis defined by $(\theta,\phi)$=(\ang{15},\ang{45}). In that case, the interfaces exhibit a mixed character which results from the competition between the interface excess energy and the periodic boundary conditions of the simulation cell. Taking advantage of \texttt{DEPMOD} could help studying the sensitivity of the critical stress surface to the strain-rate or temperature, which could have strong implications on the modeling of graphitic systems at the mesoscale, as well as microstructured carbon-based materials such as pyrocarbons~\cite{Leyssale_carbon_2022, Polewczyk_carbon_2023}. 

\subsection{\label{subsec:silicon}Silicon upon pure compression}
A single crystal of silicon with cubic diamond structure was subjected to pure compression following Equation~\ref{eq:Ftrac} along multiple directions defined using the spherical coordinates system. $\theta$ was varied between \ang{45} and \ang{90} while $\phi$ was chosen between \ang{0} and \ang{45}, both with a \ang{15} step. \texttt{DEPMOD} was used to generate the deformation paths and a total of 16 MD simulations was performed. The Si single crystal was oriented in its most general configuration, i.e. with the $\T{[100]}$, $\T{[010]}$ and $\T{[001]}$ axes along the x-, y- and z-axes.
\begin{figure}[!h]
\centering
\includegraphics[width=0.8\linewidth]{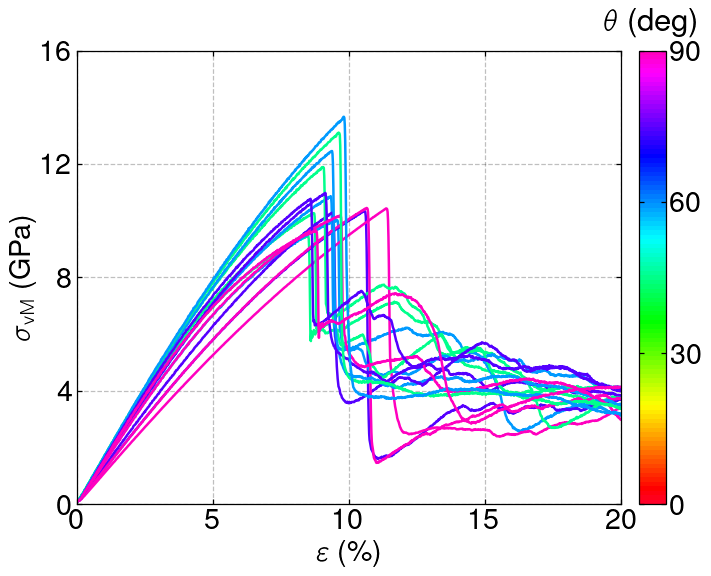}
\caption{\label{fig:curves_si} Stress-strain curves obtained for silicon single crystal under pure compression. Curves are colored according to the polar angle $\theta$ that partially defines the loading direction.}
\end{figure}
The corresponding stress-strain curves are reported in Figure~\ref{fig:curves_si} and colored according to the value of $\theta$. While the curves are not clearly grouped as for graphite, one can detect a correlation between the angle $\theta$ and the value of the critical shear stress, i.e. the maximum value of the stress at the onset of a deformation mechanism that induces an abrupt relaxation and drop of the latter. Indeed, it appears that loadings with $\theta$ close to \ang{90} (pink) have a critical stress around 10 GPa while loadings with $\theta$ close to \ang{45} degree exhibit a higher critical stress above 12 GPa. Loadings with $\theta$ in between lead to a critical stress in between as well. In addition, curves with a similar color (i.e. a similar $\theta$) appear dispersed in term of critical shear stress, indicating an additional dependence on the $\phi$ angle. Overall, the critical shear stress appears less dispersed in comparison to graphite, as seen on the reconstructed surface displayed in Figure~\ref{fig:flow_full}b) which appears more isotropic, i.e. close to a sphere. Logically, the silicon crystal symmetries are recovered in that surface which exhibits however some directional dependence of the mechanical response with respect to the loading direction.
\begin{figure}[!t]
\centering
\includegraphics[width=0.9\linewidth]{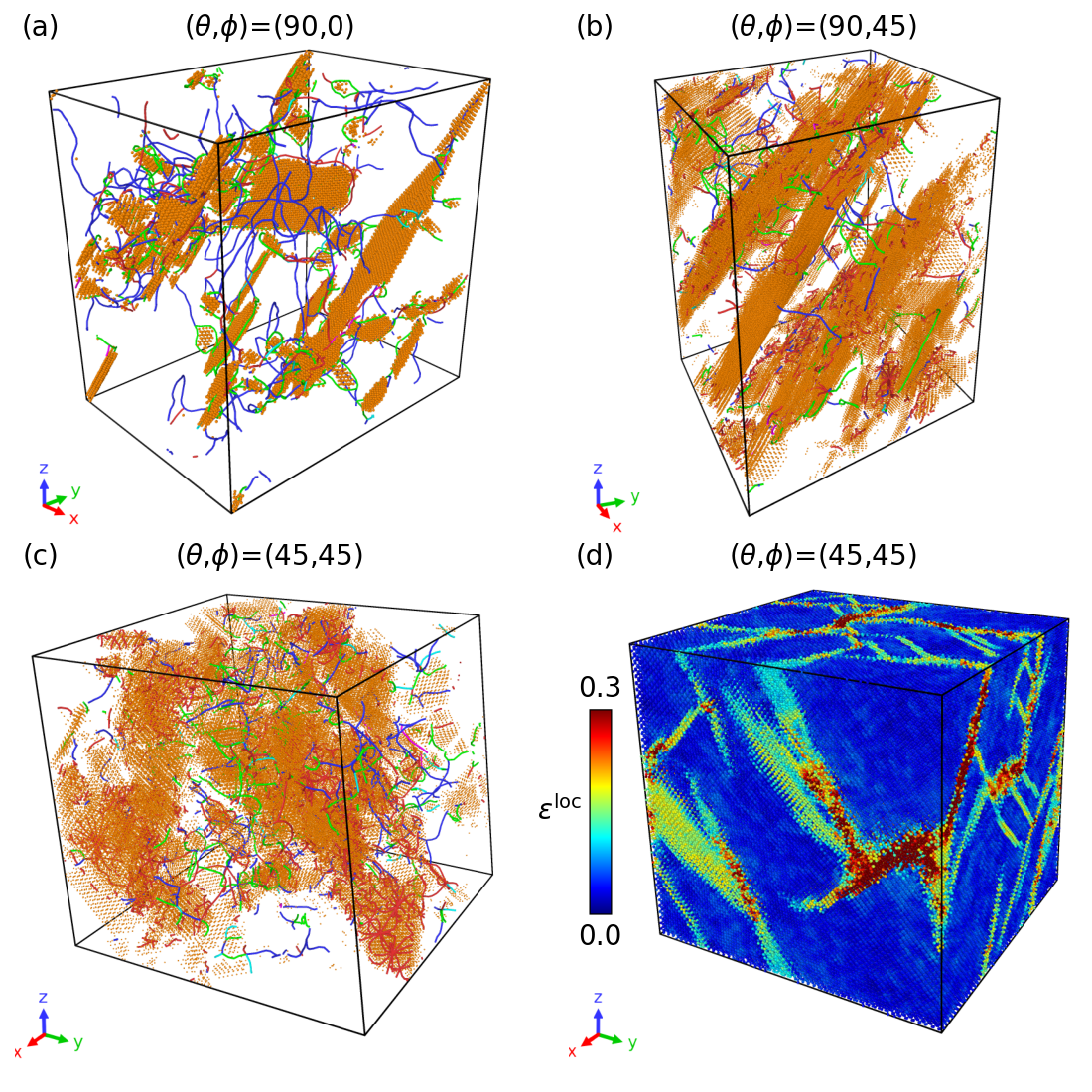}
\caption{\label{fig:ustructure_si} Microstructures observed in silicon single crystal under isochoric compression along various directions defined by $(\theta,\phi)$ in spherical coordinates. Atoms corresponding to the cubic diamond environment have been removed in (a), (b) and (c). In (d), atoms are colored according to the local value of the von Mises shear strain.}
\end{figure}
To get more insights into the differences on the critical stresses as a function of the loading direction, we display in Figure~\ref{fig:ustructure_si} snapshots of the microstructures obtained after the nucleation of the first deformation mechanisms for three different loading directions labelled \textbf{A}, \textbf{B} and \textbf{C} on Figure~\ref{fig:flow_full}. The corresponding microstructures are shown in Figure~\ref{fig:ustructure_si}a), b) and c) respectively. For direction \textbf{A}, i.e. $(\theta,\phi)$=(\ang{90},\ang{0}), the microstructure evolution is dominated by the nucleation of both perfect and partial dislocations, indicated by the presence of thin hexagonal diamond planar structures. Indeed, using the Polyhedral Template Matching~\cite{Stukowski2012} algorithm in \texttt{OVITO}~\cite{ovito_ref}, atoms with hexagonal diamond environment have been kept for visualization purposes while the remaining atoms have been removed for clarity. In addition, the Dislocation eXtraction Algorithm was applied to identify dislocation in the cubic diamond structure. Along direction \textbf{B}, i.e. $(\theta,\phi)$=(\ang{90},\ang{45}), the mechanical response of Si involves a strong plastic activity as well as the appearance of large bands of hexagonal diamond structure. While such a phase is not known for Si, it can result from partial dislocation piling. However, the microstructure is clearly different compare to the one obtained in \textbf{A}. Finally, the compression along \textbf{C}, i.e. $(\theta,\phi)$=(\ang{45},\ang{45}), led to an even more complex microstructure. On one hand, dislocations are present with a large proportion of hexagonal diamond thin plates, signature of partial dislocations in cubic diamond. On the other hand, no large and thick areas of hexagonal diamond were found in that case and the Polyhedral Template Algorithm (PTM) \cite{Stukowski2012} was not able to identify a large amount of the atoms. A recent study has discussed the appearance of various crystal phases in Si~\cite{ge2024silicon} which should be further examined in the present case, by exploiting more evolved crystal structure identification tools as previously developed~\cite{Lafourcade_cms_2023,Allera_cms_2024}. In addition, stacking faults in $\T{(111)}$ planes similar to the ones displayed in Figure~\ref{fig:ustructure_si}a) have recently been investigated using diamond anvil cell (DAC) experiment and are considered as precursors to the BC8 phase transformation. Finally, probably due to the small size of the simulation cell and the large amount of plasticity, some amorphous areas appear in the crystal, due to dislocation interactions and shear bands formation. These mechanisms are illustrated in Figure~\ref{fig:ustructure_si}d) where atoms are colored according to their local shear strain value as calculated by \texttt{OVITO}~\cite{ovito_ref}. In the representation, one can distinguish plastic events (thin cyan lines) from pseudo-phase transformation (large yellow/green areas) from amorphous shear bands (localized red bands). We believe that using \texttt{DEPMOD} along with large scale MD simulations can help better understand the mechanical response of such materials as well as discover new deformation mechanisms or phase transformation processes and their dependence on the deformation type and loading direction.

\begin{figure}[!b]
\centering
\includegraphics[width=0.8\linewidth]{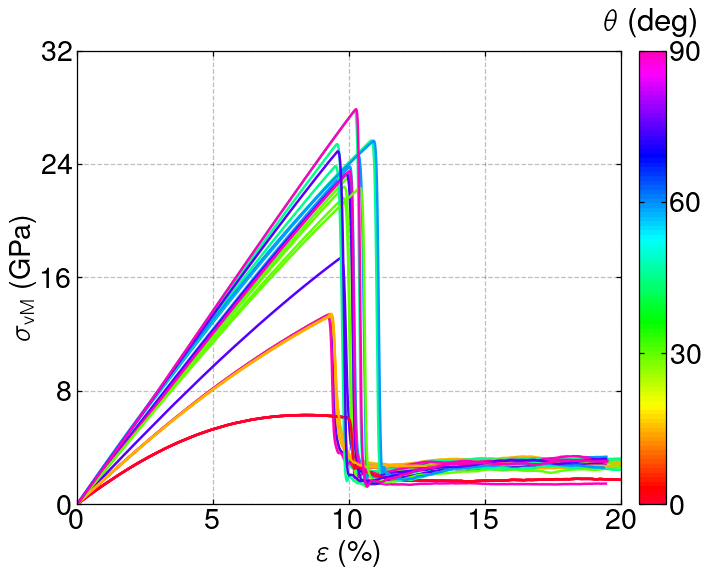} \\
\includegraphics[width=0.8\linewidth]{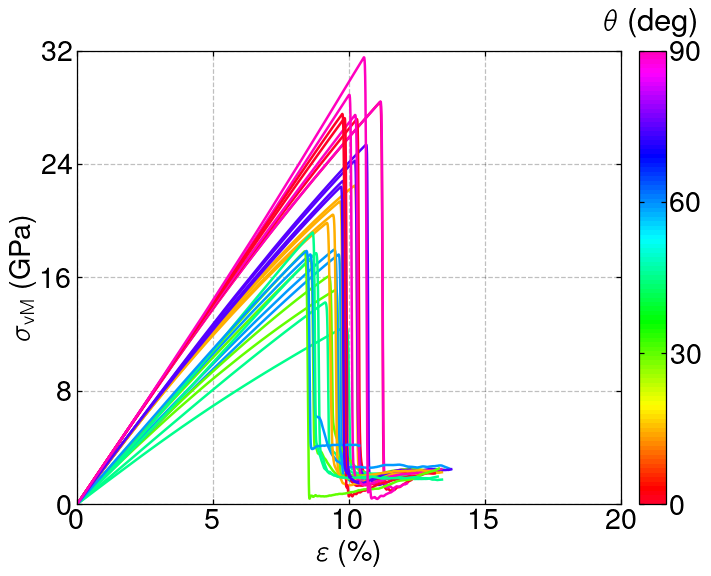}
\caption{\label{fig:curves_ta} Stress-strain curves obtained for tantalum single crystal under isochoric compression (top) and pure shear (bottom). Curves are colored according to the polar angle $\theta$ that partially defines the loading direction.}
\end{figure}

\subsection{\label{subsec:tantalum}Tantalum upon isochoric traction and pure shear}
\begin{figure}[!t]
\centering
\includegraphics[width=0.9\linewidth]{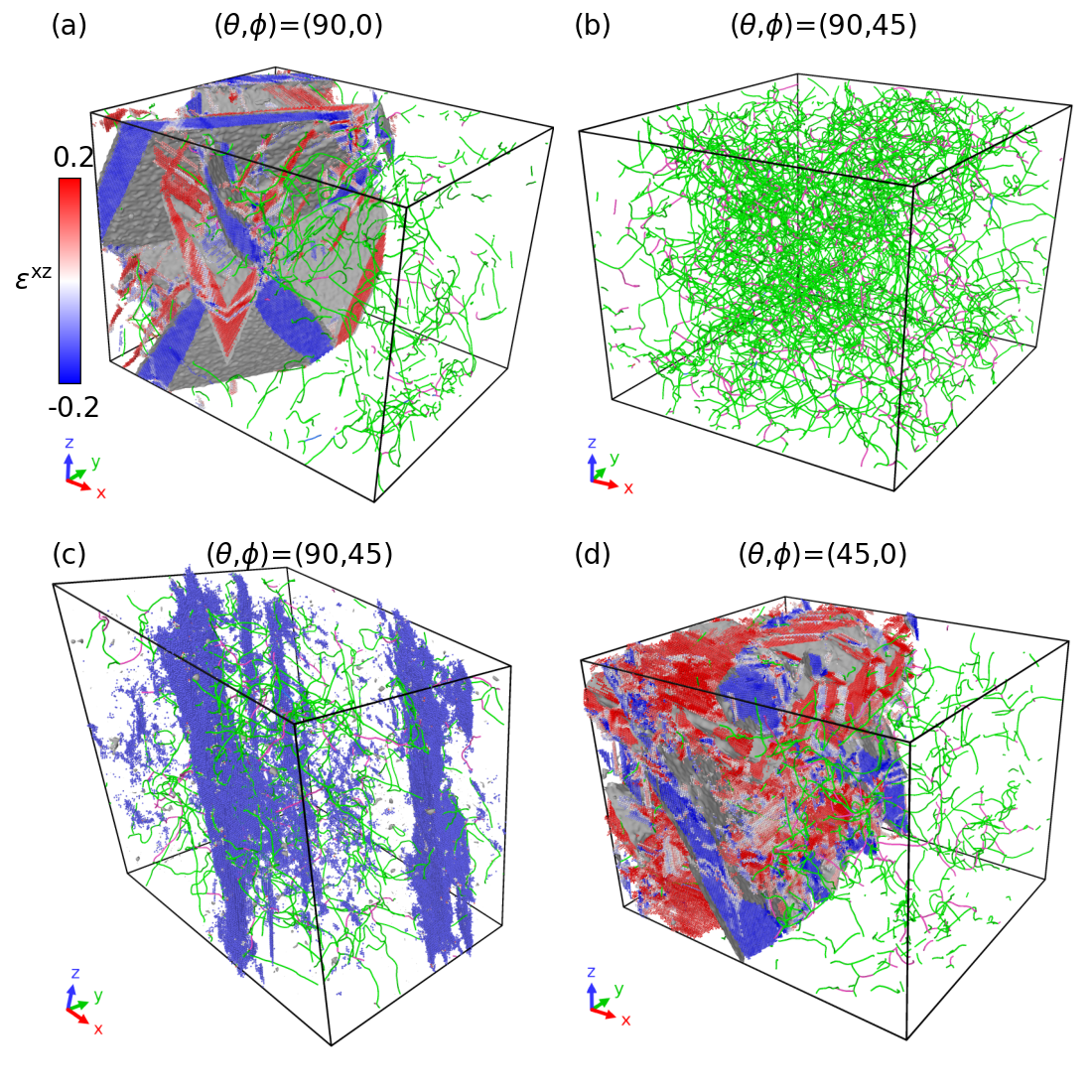}
\caption{\label{fig:ustructure_ta} Microstructures observed in tantalum single crystal under isochoric traction (a,b) and pure shear (c,d) along various directions defined by $(\theta,\phi)$ in spherical coordinates. In all snapshots, the DXA algorithm was used to identify dislocations which are represented as green and pink lines for perfect and junction dislocations respectively. Additionally, all atoms with a local shear strain lower than 0.25 have been removed. In (a) and (d), atoms are colored according to the $\varepsilon^{xz}$ component of the Green-Lagrange strain tensor, highlighting the difference in deformation orientation. In (c), atoms are colored accordingly to their local crystal structure, i.e. BCC in blue.}
\end{figure}

BCC tantalum single crystal was subjected to isochoric traction using Equation~\ref{eq:Fisovtrac}. The deformation paths were generated by specifying the loading direction using spherical coordinates through $(\theta,\phi)$ as in the previous cases. $\theta$ was varied between \ang{0} and \ang{90} while $\phi$ was chosen between \ang{0} and \ang{45}, both with a \ang{15} step.
Stress-strain curves corresponding to the isochoric traction setup are displayed in Figure~\ref{fig:curves_ta} (top) and colored according to the value of $\theta$. No specific ordering appears with respect to $\theta$ besides that all curves with $\theta \approx$ \ang{30} exhibit a similar evolution in terms of time evolution as well as shear stress level. In addition, the curves with $\theta$ = \ang{0} or \ang{90} exhibit a quasi non-linear behavior from the very beginning of the loading. In fact, these curves corresponds to a deformation path along the $\T{[100]}$ crystallographic axis that follows the Bain transformation. This transformation continuously transforms a BCC structure into a FCC one. While tantalum does not possess any FCC stable phase, this deformation path enforces it, before the apparition of a stress drop that allows BCC tantalum to reach a more stable microstructure. BCC tantalum critical shear stress surface is displayed in Figure~\ref{fig:flow_full}c). It exhibits a cubic symmetry but the critical stress value appears to strongly depend on the loading orientation. Especially, the critical stress at nucleation of the first deformation mechanism is very small for loadings along $\T{[100]}$ compared to other directions.
To better understand these differences, a snapshot of the microstructure after that stress drop (for a loading along $\T{[100]}$, labeled \textbf{A} on Figure~\ref{fig:flow_full}c) is shown in Figure~\ref{fig:ustructure_ta}a where: 1) atoms with a local shear strain lower than 0.25 have been removed, 2) atoms belonging to half part of the crystal along the x-axis have been removed and 3) dislocations have been identified using the Dislocation eXtraction Algorithm (DXA)~\cite{Stukowski2012}. BCC tantalum relaxes the stress through two main mechanisms where the first one is crystal twinning, traduced by these large deformation bands in which the local crystal structure is still BCC but with a different orientation than the parent structure, with specific orientation relations between them. The second mechanism to accommodate the applied deformation is dislocation-mediated plasticity, shown by the large amount of perfect dislocation lines (green) and a smaller amount of dislocation junctions (pink).

The microstructure obtained with a different loading direction, labeled \textbf{B} on Figure~\ref{fig:flow_full}c is displayed in Figure~\ref{fig:ustructure_ta}b and corresponds to an isochoric traction along the $\T{[110]}$ crystallographic axis. The critical stress is very large compared to the previous case as the loading does not correspond to the Bain path anymore. In the resulting microstructure, no twinning appears at all and the deformation accommodation is dominated by the nucleation of a very large amount of dislocations with a larger proportion of dislocation junctions. This highlights the complex nature of BCC tantalum mechanical response and its dependence on the loading direction. \texttt{DEPMOD} could be used in future work to study the tension-compression asymmetry of BCC tantalum as a function of loading direction, as well as strain-rate and temperature dependence in an automatic fashion. In addition, the resulting microstructures could be studied in a more exhaustive way using additional tools to quantify for example the dislocation densities evolution or the twinning volume fraction. To wrap up with the isochoric traction case, one can notice that after the stress drops, all stress-strain curves tend to a similar value around 3 GPa, except for the loading directions that contain a large twinning fraction which relax even more the stress. While before defects nucleation the curves exhibit a spread behavior, they all converge towards a single value of the flow stress which could have strong implications in modeling the plastic flow of BCC tantalum at the mesoscale~\cite{Denoual_jap_2024}. In the present work, the initial sample was considered pristine which is known to be far from the real state of a single crystalline BCC tantalum that always contains a small amount of dislocations. Future work will be dedicated to applying the present framework to samples with initial dislocations loops as done in other studies~\cite{Zepeda2017,Zepeda2020,Zeni2021}.

BCC tantalum was also subjected to directional pure shear using Equation~\ref{eq:Fshear}. The same convention as previously was used to specify the shear direction, while the shear plane was automatically determined using the procedure explained in Section~\ref{subsec:pureshear}. For clarity, a shear defined by $(\theta,\phi)$=(\ang{0},\ang{0}) corresponds to a shear direction along the z-axis in a plane with normal aligned with the x-axis. In that case, $\theta$ was varied between \ang{0} and \ang{90} while $\phi$ was chosen between \ang{0} and \ang{45}, both with a \ang{15} step. The corresponding stress-strain curves are displayed in Figure~\ref{fig:curves_ta} (bottom). In that case, All curves exhibit a similar behavior, i.e. a linear slope prior to a large stress relaxation, converging to a similar flow stress state, independent from the initial loading direction. In addition, curves appear ``sorted'' by their $\theta$ value, indicating a strong loading direction dependence. This observation correlates well with the appearance of the critical shear stress surface displayed in Figure~\ref{fig:flow_full}d. One can notice that this specific surface does not offer a cubic symmetry as for the traction case. This comes from the fact that when defining the shear deformation path, the direction $\T{m}$ is built on $(\theta,\phi)$ while the shear plane normal $\T{n}$ is built such that the plane containing $(\T{m},\T{n})$ rotates around the z-axis, considered as the pole for the spherical coordinates system. Indeed, to span all possible orientations for a shear deformation paths, one could choose any direction on the unit sphere as the pole, leading to an infinite number of possibilities. In the present work, we focus on demonstrating the versatility of the \texttt{DEPMOD} tool and leave such studies for future work. On the critical stress surface, the large values obtained for any $\phi$ and $\theta$=\ang{90} correspond to a pure shear in the plane with normal parallel to the z-axis. The resulting microstructure for a shear direction along the $\T{[110]}$, labelled \textbf{C} on Figure~\ref{fig:flow_full}d, is displayed in Figure~\ref{fig:ustructure_ta}c. Atoms with a local shear strain lower than 0.25 have been removed while the remaining ones are colored according to their crystal structure where blue means BCC. The large bands that formed in the microstructure indicate that localized shear happened, probably induced by multiple dislocation glide on these planes across periodic boundary conditions. However, no structural change is observed and these bands are pure BCC with no misorientation meaning that they do not correspond to twins. In addition, the DXA algorithm reveals a large amount of dislocations with a small proportion of junctions. Another direction labeled \textbf{D}, corresponding to a pure shear along the $\T{[101]}$ direction in a plane with normal $\T{[10\bar{1}]}$ is considered as an example for its very low critical stress compared to direction \textbf{C}. The resulting microstructure is displayed in Figure~\ref{fig:ustructure_ta}d. Similarly to Figure~\ref{fig:ustructure_ta}a, half the atoms have been removed and only atoms with a local shear strain greater than 0.25 are displayed on the other half. In addition, dislocations lines are displayed and prove to be present in a proportion similar to case \textbf{A} and \textbf{C}. In that case, the large amount of sheared material proves to be very thin planes, signature of large plastic activities. No large twins were found in the microstructure, indicating that for this specific direction, a large amount of slip systems may be activated upon pure shear. Overall, the behavior of BCC tantalum for the pure shear case appear quite different than the traction case where a competition between twinning and plasticity could be at play~\cite{Bruzy_jmps_2022}. Applying pure shear deformation paths on system with pre-existing dislocations seeded in the crystal could help understand the competition between slip systems and the role of twinning in the mechanical response of BCC tantalum. We believe that the present framework and the large variety of scenarios that offers \texttt{DEPMOD} could help to better understand the mechanical response of a wide range of materials with different properties, as demonstrated in this work.

\section{\label{sec:ccls}Conclusions and perspectives}
Building on the legacy left by Rahman and his pioneering work on materials deformations, we present a novel method for constructing deformation paths in molecular dynamics (MD) simulations, addressing the limitations imposed by existing MD engines like \texttt{LAMMPS}. By introducing the DEformation Paths for MOlecular Dynamics (\texttt{DEPMOD}) software, we enable the exploration of critical flow stress surfaces of materials under an infinite possibility of loading directions. 
We demonstrate \texttt{DEPMOD} capabilities by applying isochoric traction, pure compression, and pure shear to graphite, silicon and tantalum pristine single crystal respectively at constant strain-rate. The diversity of observed mechanical responses is, to our opinion, a direct highlight of the anisotropic nature of material's behavior and provides deep insights into the deformation mechanisms activated by directional loading. 
This approach not only advances our understanding of material mechanical response but also offers a powerful framework for integrating atomistic-level insights into multiscale models. Future developments of this work may further enhance the versatility of \texttt{DEPMOD} and broaden its application to more complex deformation processes. In addition, its application to defective materials (dislocations) might be of crucial importance to distinguish pristine and defective materials mechanics but also to identify microstructures/properties relationship.

\section*{Data availability}
  The DEPMOD code as well as corresponding tutorials and examples will be made available on \href{https://github.com/lafourcadep/depmod/}{GitHub (https://github.com/lafourcadep/depmod/)}.


\end{document}